\documentclass[aps,pra,twocolumn,superscriptaddress,a4paper,10pt]{revtex4-2}
\usepackage[T1]{fontenc}
\usepackage[utf8]{inputenc}
\usepackage{array}
\usepackage{multirow}
\usepackage{graphicx}
\usepackage{amsmath,amsfonts,amssymb,color}
\usepackage{amsthm}
\usepackage{mathtools}
\usepackage{leftidx}
\usepackage{xcolor}
\usepackage{dcolumn}
\usepackage{bm}
\usepackage{epstopdf}
\usepackage{epsfig}
\usepackage{environ}
\usepackage{pdfcomment}
\usepackage{float}
\usepackage{tabularray}
\usepackage{hyperref}
\usepackage{upgreek}

\usepackage{esint}
\usepackage{physics}
\usepackage[english]{babel}

\hypersetup{colorlinks=true,citecolor=blue,
	linkcolor=blue,urlcolor=blue,pdfstartview=FitH,
	bookmarksopen=true}
\usepackage{cleveref}

\begin{document}
\title{Unitary Transformations using Robust Optimal Control on a Cold Atom Qudit}

\author{E. Flament}
\affiliation{Laboratoire Collisions, Agr\'egats, R\'eactivit\'e, FeRMI, Université  de Toulouse, CNRS, 118 Route de Narbonne, F-31062 Toulouse Cedex 09, France}
\author{N. Ombredane}
\affiliation{Laboratoire Collisions, Agr\'egats, R\'eactivit\'e, FeRMI, Université  de Toulouse, CNRS, 118 Route de Narbonne, F-31062 Toulouse Cedex 09, France}
\author{F. Arrouas}
\affiliation{Laboratoire Collisions, Agr\'egats, R\'eactivit\'e, FeRMI, Université  de Toulouse, CNRS, 118 Route de Narbonne, F-31062 Toulouse Cedex 09, France}
\author{D. Ronco}
\affiliation{Laboratoire Collisions, Agr\'egats, R\'eactivit\'e, FeRMI, Université  de Toulouse, CNRS, 118 Route de Narbonne, F-31062 Toulouse Cedex 09, France}
\author{B. Peaudecerf}
\affiliation{Laboratoire Collisions, Agr\'egats, R\'eactivit\'e, FeRMI, Université  de Toulouse, CNRS, 118 Route de Narbonne, F-31062 Toulouse Cedex 09, France}
\author{D. Sugny}
\affiliation{Laboratoire Interdisciplinaire Carnot de Bourgogne, CNRS UMR 6303, Universit\'{e} de Bourgogne, BP 47870, F-21078 Dijon, France}
\author{D. Gu\'ery-Odelin}
\affiliation{Laboratoire Collisions, Agr\'egats, R\'eactivit\'e, FeRMI, Université  de Toulouse, CNRS, 118 Route de Narbonne, F-31062 Toulouse Cedex 09, France}

\date{\today}

\begin{abstract}
In this paper, we design and experimentally implement various robust quantum unitary transformations (gates) acting on $d$-dimensional vectors (qudits) by tuning a single control parameter using optimal control theory.
The quantum state is represented by the momentum components of a Bose-Einstein condensate (BEC) placed in an optical lattice, with the lattice position varying over a fixed duration serving as the control parameter. To evaluate the quality of these transformations, we employ standard quantum process tomography. In addition, we show how controlled unitary transformations can be used to extend state stabilization to global stabilization within a controlled vector subspace.
Finally, we apply them to state tomography, showing how the information about the relative phase between distant momentum components can be extracted by inducing an interference process.
\end{abstract}

\maketitle

\section{Introduction}
\label{sec:level1}
Quantum technologies have garnered growing interest over the past decades, as they hold the potential to revolutionize information processing, communications, and metrology, surpassing the capabilities of current technologies \cite{69f6dd0d82914a7585c1cdf2aa05681b, Cirac:2012huh}.

 To advance these technologies, numerous quantum platforms are currently under investigation: trapped ions \cite{PhysRevLett.130.050803, HAFFNER2008155,PhysRevX.8.031022,ioncomp}, superconducting circuits \cite{devoret2013superconducting,superconnature,superconscience,KIKTENKO20151409, Goss_2022}, quantum dots \cite{kloeffel2013prospects}, cold atoms \cite{PhysRevA.104.L060601, PRXQuantum.2.040303, PhysRevLett.131.240801}, optical cavities \cite{Ralph_2010} to name just a few. Most of these technologies operate within the paradigm of binary information theory, utilizing qubits. However, it has been demonstrated that qudits—the $d$-dimensional generalization of qubits—enable faster and more reliable information processing. Compared to qubits, qudits offer advantages such as reduced quantum circuit depth, a more natural mapping for many-body simulations, and more efficient error correction schemes \cite{PhysRevLett.114.240401, rambow2022reductioncircuitdepthmapping, PhysRevX.13.021028,PRXQuantum.4.040333, 10.3389/fphy.2020.589504, luo2014universal, Cao_2024, Jankovi__2024, vezvaee2024quantumsimulationfermihubbardmodel,PhysRevLett.129.160501, PhysRevLett.113.230501}. 

Beyond the size of the logical unit, the reliability of quantum technologies hinges on the precise manipulation of quantum systems—whether to prepare specific quantum states or implement quantum processes. A widely applicable and highly effective strategy for these tasks is the use of optimal control theory \cite{koch2022quantum}. Optimal control has been applied successfully to state preparation accross various quantum platforms \cite{PRXQuantum.2.040303, Rodzinka:2024gtf,PhysRevResearch.7.013246, PhysRevA.90.023824, PhysRevX.10.021058, PhysRevLett.118.150503}. It has also shown promising results in performing quantum processes \cite{PhysRevLett.114.200502, PhysRevApplied.19.034068, PhysRevA.104.L060601, PhysRevLett.111.170502, PhysRevLett.114.240401, PhysRevLett.89.188301, Cao_2024}.
Under controllability conditions, optimal control protocols can be employed to design any unitary transformation, and consequently, any universal gate set, providing exceptional versatility \cite{PhysRevResearch.7.013246}. Moreover, these protocols can be tailored to withstand internal and external sources of error, enhancing their robustness \cite{PhysRevLett.132.193801, kosut2022robustquantumcontrolanalysis, Rodzinka:2024gtf}, and offering a reliable tool in the NISQ era. 
 
In this work, we apply the optimal control approach to unitary transformations of a cold-atom qudit. We enhance the control design by formulating the algorithms in Fourier space in order to address experimental constraints and ensure strong robustness against parameter fluctuations using the averaging method \cite{kosut2022robustquantumcontrolanalysis, WEIDNER2025111987}.
In Sec.~\ref{section2},  we present the methodology for implementing such robust quantum unitary processes on a vector composed of the discretized momentum components of a Bose-Einstein condensate (BEC) confined in a one-dimensional optical lattice, with the lattice position serving as the single control parameter. In Sec.~\ref{section3}, we describe the characterization of the experimental unitary transformations using standard quantum process tomography \cite{nielsen2010quantum}. Section \ref{section4} presents our experimental results and benchmarks of various processes, including four qubit gates and a qutrit gate, offering a novel approach to quantum computing with atom qudits. In Sec.~\ref{section5}, we discuss applications of these protocols, first by demonstrating global stabilization of a Hilbert subspace, building upon concepts from state-to-state control protocols and Floquet engineering \cite{CRPHYS_2023__24_S3_173_0}. Finally, we illustrate how these unitary processes can be applied to enhance state tomography capabilities in combination with existing techniques.

\section{\label{sec:exp_alg}Experimental setup and algorithm}
\label{section2}

In this section, we describe the quantum system under study, the design of robust, custom-tailored optimal control strategies, and the measurements we perform to extract information from the system. 

\subsection{\label{exp}BEC experimental setup}

Our experimental setup produces a pure Bose-Einstein condensate (BEC) of typically $5\times10^5$ atoms of $^{87}$Rb in the $\ket{F=1,m_F=-1}$ hyperfine state in a hybrid trap made of a crossed optical dipole trap in the horizontal plane superimposed on  a quadrupolar magnetic trap \cite{PhysRevLett.117.010401}. The optical lattice, aligned with one of the dipole beams, is generated by the interference of two counter-propagating laser beams with a wavelength of $\lambda=1064\,$nm. By calling $x$ the axis of the optical lattice, the BEC experiences the following potential: 
\begin{equation}
    \hat{V}(\hat{x},t)=-\frac{s E_L}{2}\cos(k_L \hat{x}+\phi(t))+\hat{U}_{hyb}(\hat{x}, \hat{y}, \hat{z}).
\end{equation}
Here, $k_L \equiv 4\pi/\lambda$ is  the optical lattice wave number and $E_L=\hbar^2 k_L^2 /(2m)$ (with $m$ the atomic mass) serves as a natural energy scale for the lattice. The dimensionless lattice depth $s$ can be precisely calibrated \cite{PhysRevA.97.043617} with a typical uncertainty of $3\%$. The frequency of the three-dimensional harmonic hybrid trap along the $x$ axis is $\omega_x \simeq 2\pi \times 7\,$Hz. Since our experiments operate on a sub-millisecond timescale, the contribution of this trapping potential to the dynamics can be neglected when designing the optimal control algorithm.

The BEC is initially adiabatically loaded into the optical lattice, ensuring that the quantum state corresponds to the lowest energy Bloch state in the band structure of the periodic potential~\cite{Chatelain_2020}. This process places the BEC to a good approximation in the subspace of zero quasi-momentum, a quantity which is conserved throughout the dynamics. We can express the quantum state of the BEC in the momentum basis as:
\begin{equation}
    \ket{\Psi}=\sum_{\ell \in \mathbb{Z}}c_\ell \ket{\chi_\ell},
    \label{latice_quantum_state}
\end{equation}
where $\chi_{\ell}(x)=e^{i l k_L x}/\sqrt{\lambda/2}$.

To extract information about the BEC, we abruptly switch off all traps and allow the BEC to expand ballistically for approximately 35 ms. An absorption image is then taken, providing direct access to the populations $|c_\ell|^2$ in each diffraction order (i.e. momentum components). 

The lattice amplitude $s$ and phase $\phi(t)$ are finely tuned by the amplitude and relative phase between the drives of two  phase-locked acousto-optic modulators (AOMs) placed in the paths of the lattice beams. Optimal control protocols can then be implemented to prepare arbitrary states~\cite{PRXQuantum.2.040303}.

\subsection{\label{alg}Robust Fourier optimal control algorithm}

\noindent\textit{Introduction}. In order to implement controlled unitary transformations, we use $\phi(t)$ as the single control parameter over a fixed control duration $t_f$. In the momentum basis, Schrödinger's equation for the dynamics translates into a set of coupled linear equations:
\begin{equation}
    i\dot{c}_\ell=\ell^2c_\ell-\frac{s}{4}(e^{i\phi(t)}c_{\ell-1}+e^{-i\phi(t)}c_{\ell+1}),
\end{equation}
 where the time has been rescaled as  $t\rightarrow E_L t/\hbar$. We consider unitary transformations in a finite-dimensional restriction of the infinite-dimensional Hilbert space. For qudits made of $d$ momentum components, this operation can be represented by the action of the operator $\hat{U}(t_f, \phi)$ which maps one basis of the qudit Hilbert space to another basis of the same space. More precisely, this operator can be expressed as $\hat{U}(t_f, \phi) =\ket{C_{n_1}'}\bra{C_{n_1}}+...+\ket{C_{n_d}'}\bra{C_{n_d}}$, where $t_f$ is the duration of the transformation, and $\{\ket{C_n}\}$ (resp. $\{\ket{C'_n}\}$) are two basis of the qudit Hilbert space, with components $\{C_{n,\ell}\}$ (resp. $\{C'_{n,\ell}\}$), see Eq.~\eqref{latice_quantum_state}.

In the following, we outline the optimal control algorithm used to design $\phi(t)$.
\\

\noindent\textit{Optimal control algorithm in Fourier space}.
In practice, we use a first order gradient ascent algorithm \cite{ PRXQuantum.2.040303, PhysRevLett.89.188301, Werschnik_2007} that maximizes the matrix fidelity:
\begin{equation}
F\equiv {\bigg|\bra{\hat{U}(t_f, \phi)}\ket{\hat{U}_{T}}\bigg|}^{2}\equiv \frac{1}{d^{2}}|\mbox{tr}(\hat{U}^{\dag}_{T}\hat{U}(t_f, \phi))|^{2},
\label{Unitary_fidelity}
\end{equation}
where $\hat{U}(t_f, \phi)$ is the designed unitary operator obtained from the control algorithm through the varying phase $\phi$, $\hat{U}_{T}$ the targeted unitary restricted to the subspace of size $d$.

A natural timescale for the dynamics from the lattice's ground state is given by $T_0=\hbar/\Delta E$ where $\Delta E$ is the energy difference between the two lowest eigenstates at zero quasi-momentum. For a typical lattice depth used in the experiments described below, $s_0\approx 5$, we find $T_0\approx 60$ $\upmu$s ($1/T_0=16.67$ kHz) \cite{PRXQuantum.2.040303}, which sets a typical timescale for the control duration $t_f$. 
Another constraint on the control arises from the finite bandwidth of the modulation system, which limits the frequency range available for the control. To accommodate these different constraints, we expand the control ramp as a Fourier series
\begin{equation}
\phi(t)=a_0+\sum_{n=1}^{n_{max}}(a_n \cos(2\pi n f_0 t)+
b_n \sin(2\pi n f_0 t)),
\end{equation}
with $f_0=1/t_f$
the frequency associated to the optimal control duration and $n_{max}$ the number of harmonics used to design the control protocol. For our experiments, we set $f_{max}= 125$ kHz, which defines $n_{max}=\lceil\frac{f_{max}}{f_0}\rceil$. This allows fast phase variations while staying safely below the cutoff frequency of the AOMs. 

\begin{figure}
\centerline{
    \includegraphics[scale=0.55]{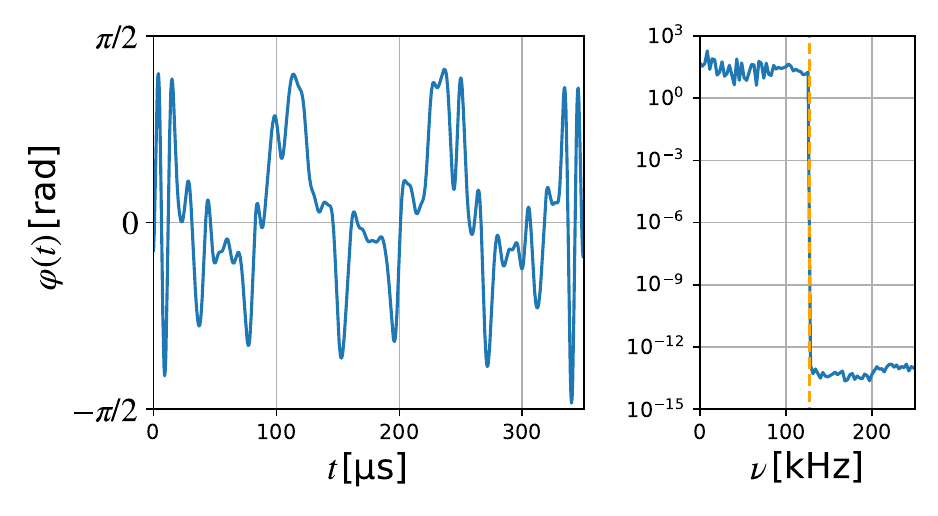}}
    \caption{Left panel: Phase control protocol after numerical optimization in Fourier space for an $\hat{X}$ gate, over a time interval of $350$ $\upmu$s. Right panel: Absolute value of the FFT of the phase control signal shown on a logarithmic scale. The frequency cutoff at $f_c= 125$ kHz is indicated by the orange dashed line (see \ref{alg}).}
    \label{rampe_fourier_X}
\end{figure}

The control duration is bounded from below for a given lattice depth by the quantum speed limit~\cite{QSLreview}. This bound is expected to be larger than in the case of a state-to-state transformation, as the unitary control procedure here applies to a set of states. 
Furthermore, the frequency restriction along with robustness requirements (see below) impose strong additional constraints on the control. 
As a result, we find here a good convergence of the control algorithms for final times $t_f$ ranging from $300$ $\upmu$s to $450$ $\upmu$s, \emph{i.e.} 5 to 8 $T_0$ whereas it was only 1.5$T_0$ for unconstrained state-to-state transformation \cite{PRXQuantum.2.040303}. Figure \ref{rampe_fourier_X} provides an example of a typical phase control $\phi(t)$ along with its frequency spectrum.

To design the control unitary operator, we assume
that the control $\phi$ is a piecewise constant function in time
with a time step $\Delta t=t_f/k$, where $k$ is the number of time steps. The evolution operator can be expressed as
\begin{equation}
    \hat{U}_f= \hat{U}_{k}\hat{U}_{k-1}...\hat{U}_1, 
\end{equation}

\noindent with $\hat{U}_j=\exp(-i \Delta t \hat{H}(\phi(j\Delta t)))$ and $\hat{U}_f\equiv\hat{U}(t_f, \phi)$. The approximation holds well for small enough time steps. In our experiments, we can safely use $\Delta t = 500$ ns. 

From now, we will use the notation $\phi(j\Delta t)\equiv \phi_j$ and $j\Delta t\equiv t_j$. The control protocol in Fourier space is numerically found by maximizing the fidelity $F$ with respect to the Fourier coefficients $a_n$ and $b_n$ through the values of each $\phi_j$ using an iterative algorithm. This requires the calculation of the partial derivative of $F$ with respect to the coefficients $\{a_n,b_n\}$ at each time step: 
\begin{align*}
\begin{split}
\frac{\partial F}{\partial a_n }\bigg\vert_{t_j}
&=2 \mathcal{R}e\bigg[ \frac{\partial \hat{U}_{f}^{\dag}}{\partial \phi_j}\frac{\partial F}{\partial \hat{U}_{f}^{\dag}}\bigg]\frac{\partial \phi_j}{\partial a_n} \approx 2\Delta t \frac{\partial \phi_j}{\partial a_n} \times \\
\end{split}
\\
&\mathcal{I}m \bigg[\bra{\hat{U}_k \hat{U}_{k-1}...\hat{U}_{k+j}\frac{\partial \hat{H}_j }{\partial \phi_j}\hat{U}_{k-j}...\hat{U}_1}
\ket{\hat{U}_{T}}\bra{\hat{U}_{T}}\ket{\hat{U}_{f}} \bigg].
\end{align*}

The total gradient reads $\frac{\partial F}{\partial a_n}=\sum_j \frac{\partial F}{\partial a_n }\big\vert_{t_j}=\sum_j \frac{\partial F}{\partial \phi_j}\frac{\partial \phi_j}{\partial a_n}$. A similar expression can be derived for the gradient with respect to $b_n$.
Alternatively, this derivation can be obtained from the Pontryagin's principle \cite{Ansel_2024, PRXQuantum.2.030203, PRXQuantum.2.040303}. In practice, the numerical algorithm proceeds as follows:
\begin{enumerate}
\item Select an initial guess for the Fourier coefficients  $\{a_n; b_n\}_0$.
\item Compute the final unitary $\hat{U}(t_f, \phi)$ for this set of coefficients.
\item Compute the gradients $\partial F/\partial a_n$ and $\partial F/\partial  b_n$.
\item Update each coefficient by applying the gradient correction weighted by a small positive parameter $\varepsilon$: $a_n \leftarrow a_n+\varepsilon \partial F/\partial a_n$, $b_n \leftarrow b_n+\varepsilon \partial F/\partial b_n$.
The $\epsilon$ coefficient is dynamically adjusted  to improve algorithm convergence and/or to avoid local minima.
\item Repeat from step 2 until a stopping condition is met, such as a sufficiently small correction factor $\epsilon$ or an acceptable level of fidelity.
\end{enumerate}

\noindent\textit{Improved algorithm to ensure robustness}. 
We used the concurrent averaging method \cite{Dionis_2022}\footnote{Although the sequential averaging method should perform better in terms of convergence rate, the concurrent averaging allows for parallelized hence faster computation of the controls.}, to tailor protocols robust against lattice depth variations or uncertainties. For this purpose, we introduce an average fidelity over $N_s$ lattice depths chosen in a given interval $[s_1, s_2]$ (we use $3\leq N_s\leq 8$): 
\begin{equation}
   \bar F =  \frac{1}{N_s}\sum_{s_0 \in [s_1, s_2]}F(s_0), 
\end{equation}
where $N_s$ is the number of terms in this sum. The method involves applying the first three steps of the algorithm detailed in the previous section for each sampled value of the lattice depth, with a modification to the fourth step. The only change is to average the corrections over the sampled lattice depth values before applying them. Interestingly, this approach only requires minimal changes to our algorithm.

\section{Quantum process certification}
\label{section3}

In this section, we review the formalism used for standard quantum process tomography (SQPT), outline the procedure for its implementation, and define the two fidelity measures used to benchmark the reconstructed process against the target unitary operator.\\ 

\noindent\textit{Quantum process}. A general quantum process reads:
\begin{equation}
    \epsilon(\rho)=\sum_{i}A_i \rho A_i^{\dagger},
\end{equation}
where $A_i=\sum_{m}a_{im}E_m$ are Kraus's operators respecting $\sum_i A^{\dagger}_i A_i\leq \mathbb{I}_N$, and the ($E_m$) form a complete basis of the space of operators on the finite-dimensional Hilbert space. In our context, $\rho$ represents the initial density matrix, and $\epsilon(\rho)$ the density matrix after the time evolution governed by the optimal control. The inequality describes the possible loss of probability from the qudit subspace, as momentum components outside it may be populated. This can be due to a residual difference between the derived control and the target process, or experimental imperfections that cause. 
the phase variation experienced by the atoms to deviate from the intended design.

It is useful to characterize the quantum process under study 
by introducing the process matrix (also known as the chi-matrix):
\begin{equation}
    \epsilon(\rho)=\sum_{mn}\chi_{mn} E_m \rho E_n^{\dagger}.
\end{equation}
Standard quantum process tomography (SQPT) refers to the procedure of retrieving the elements of the process matrix. Note that the parameters $\chi_{mn}$ do not depend on the density operator $\rho$.\\

\noindent\textit{Fidelities}. The literature introduces two fidelities to compare matrix processes. We provide here their definitions along with their relation. To compare two processes, $\epsilon_a$ and $\epsilon_b$, some authors use the so-called process fidelity defined as \footnote{This formula can be derived from the entanglement fidelity used for quantum chanels (see Ref.~\cite{NIELSEN2002249}).} \cite{PhysRevResearch.3.033031, PhysRevA.82.042307}:
\begin{eqnarray}
    F_p(\epsilon_{a}, \epsilon_{b})&= & \frac{1}{\mathcal{N}}\sum_{m}\mbox{tr}\left[\epsilon_a(E_m) \epsilon_b(E_m)^\dagger\right],\nonumber \\ & =&\frac{1}{\mathcal{N}}\mbox{tr}\left[\chi^{(a)}\chi^{\dagger (b)}\right].
    \label{process_fidelity}
\end{eqnarray}
It is nothing but the mean fidelity between the two considered processes of each 
operator $E_m$. 

In our experiments, we consider only unitary targets $\hat{U}_T$, that we compare to experimental processes, reconstructed using SQPT (see below). We choose the normalization factor~\footnote{The normalization factor $\mathcal{N}=\sqrt{\mbox{tr}(\chi^{\dagger(a)}\chi^{(a)})\mbox{tr}(\chi^{\dagger(b)}\chi^{(b)})}$ is commonly used in the presence of losses \cite{PhysRevA.82.042307, WANG200858} which is relevant when the losses are undetected events and one wants to not take into account these events. This is not the case in our experiment, since the losses are measured as populated momentum components outside the controlled subspace.} to be $\mathcal{N}=d^2$.

Another useful metric is the average gate fidelity. If we compare an experimental process $\epsilon_{exp}$ to a target unitary process $\epsilon_{\hat{U}_T}$, it is defined as: 
\begin{equation}
    F_{avg}(\epsilon_{\hat{U}_T}, \epsilon_{exp}), =\int d\psi_i \   \mbox{tr}\left[\hat{U}_T\ket{\psi_i}\bra{\psi_i}\hat{U}^{\dagger}_T \epsilon_{exp}(\ket{\psi_i}\bra{\psi_i}) \right],
    \label{average_gate_fidelity}
\end{equation}
where the integral is taken over all the pure input states. It is the average state fidelity between the initial state $| \psi_i \rangle \langle \psi_i |$ transformed by the target unitary operator $\hat{U}_T$ and the experimental process $\epsilon_{exp}(\psi_i)$.

Interestingly, for non-trace-preserving processes (arising when momentum components outside the qudit subset become contaminated), the two fidelities introduced above are directly related by the following expression (see Appendix \ref{appendix}):
\begin{equation}
    F_{avg}(\epsilon, \epsilon_{\hat{U}})=\frac{dF_p(\epsilon, \epsilon_{\hat{U}}) + \alpha_{avg}}{d+1}, 
    \label{general_fid_formula}
\end{equation}
with
\begin{equation}
    \alpha_{avg}=\frac{1}{d}\mbox{tr}\left[\epsilon(\mathbb{I}_d)\right]=\frac{1}{d}\sum_i \mbox{tr}\left[\epsilon(\ket{i}\bra{i})\right].
\end{equation}
The coefficient $\alpha_{avg}$ is straightforward to evaluate, as it represents the average population transferred to the desired qudit subspace after applying the process. For trace-preserving processes, this formula simplifies to the one defined in Ref.~\cite{NIELSEN2002249}. \\

\noindent\textit{Canonical basis}.
In this work, we adopt the SQPT conventions by using the canonical basis for complex $d \times d$ matrices, to express quantum processes and the associated fidelities \cite{nielsen2010quantum}. 
Consider the following matrix basis $\mathcal{B}=\{\left(B_{uv}\right)_{i,j}=\delta_{ui}\delta_{vj}; (u,v)\in [0,d-1]^2\}$ with $d$ the size of the associated Hilbert space $\mathcal{H}=\{\ket{u}, u \in [0,d-1]\}$, with the properties $\mbox{tr}[B_{uv}B_{kl}]=\delta_{ul}\delta_{vk}$.
The effect of a given quantum process on a basis element can be expressed as:
\begin{equation}
\epsilon(B_{uv})=\sum_{kl}\beta_{uvkl} B_{kl},
\end{equation}
As in Eq.~\eqref{process_fidelity}, the process fidelity becomes :

\begin{align}
F_p(\epsilon_{a}, \epsilon_{b})&=\frac{1}{\mathcal{N}}\sum_{uv}\mbox{tr}\left[\epsilon_a(B_{uv})\epsilon_b(B_{uv})^{\dagger}\right],\nonumber\\
&=\frac{1}{\mathcal{N}}\sum_{uv}\sum_{kl} \beta_{uvkl}^{(a)}\bar{\beta}_{uvkl}^{(b)},\nonumber\\
&=\frac{1}{\mathcal{N}}\mbox{tr}\left[\underline{\underline{\beta^{(a)}}}\,\underline{\underline{\beta^{\dagger (b)}}}\right],
\label{fid_2}
\end{align}
with $\underline{\underline{\beta^{(a)}}}=\sum_{uv}\sum_{kl} B_{uv} \otimes \beta_{uvkl}^{(a)}B_{kl}\equiv \sum_{uv} B_{uv} \otimes \epsilon_a(B_{uv})$
the new super-operator that encapsulates the quantum process $\epsilon_a$ (likewise for $\underline{\underline{\beta^{(b)}}}$). This is referred to as the Choi matrix associated with a quantum process. This fidelity formula is similar to Eq.~\eqref{process_fidelity}, here expressed in the canonical basis $\mathcal{B}$~\cite{Johnston_2011}.\\

\noindent\textit{Retrieving the super-operator elements.\label{SQPT_method}} 
Standard quantum process tomography provides a method to extract the matrix elements of the $\underline{\underline{\beta}}$ matrix. To achieve this, we must determine the effects of the processes associated with the elements of the basis $\mathcal{B}$. 
Since $B_{uv}$ only represents a physical state for $u=v$, the quantum process for $B_{uv}$ with $u\neq v$ is reconstructed by combining the results of the process applied to physically accessible states:

\begin{eqnarray}
\epsilon(B_{uv})=&&\epsilon(\ket{+}\bra{+}_{uv})-
i\epsilon(\ket{-}\bra{-}_{uv}) \nonumber\\
&&-\frac{(1-i)}{2}(\epsilon(B_{uu})+\epsilon(B_{vv})),
\end{eqnarray}

\noindent with:
\begin{align}
    \ket{+}\bra{+}_{uv}&=\frac{1}{2}(\ket{u}+\ket{v})(\bra{u}+\bra{v}), \\
    \ket{-}\bra{-}_{uv}&=\frac{1}{2}(\ket{u}+i\ket{v})(\bra{u}-i\bra{v}).
\end{align}
This equation only involves physically valid input states. Consequently, complete process tomography requires performing state tomography on $d^2$ input states. State tomography is carried out using a maximum likelihood algorithm~\cite{dupont2023phase}, which yields a density matrix $\rho$ with purity $P=\mbox{tr}\left[\rho^2\right]$. The reconstructed state $\rho$ is compared to a target state $\rho_T$ using the state fidelity $F_s=\mbox{tr}\left[\sqrt{\sqrt{\rho_T}\rho\sqrt{\rho_T}}\right]^2$. In this way, the experimental process can be compared to the target using Eqs.~\eqref{average_gate_fidelity} and \eqref{fid_2}. It is to be reminded that the state tomography yields a density matrix in an extended basis state $\mathcal{H'}$, larger than the computational basis $\mathcal{H}$. 

It is important to note that SQPT assumes perfect input states, which is rarely the case in experimental devices.  However, this problem can be mitigated using robust quantum protocols for both the quantum processes and the initial state preparation as discussed in Sec.~\ref{alg}. 

\section{Experimental gates}
\label{section4}

In this section, we demonstrate the experimental realization of gates acting on Hilbert spaces of dimension 
$d=2$ and $d=3$, using the previously mentioned maximum likelihood estimation algorithm to perform the state tomography required for SQPT. 
More specifically, we performed SQPT on the three Pauli gates $\hat{X}, \hat{Y}, \hat{Z}$ and the Hadamard gate ($\hat{H}$) for qubit gates associated with the symmetrized momentum subspace $\mathcal{H}=\{\ket{-1}, \ket{1}\}$. We also conducted SQPT on the qutrit gate $X^{(12)}$ within the momentum subspace $\mathcal{H}=\{\ket{-1}, \ket{0}, \ket{1}\}$.

\subsection{Qubit gates \label{qubit_gate}}

\begin{figure}[h!]
    \includegraphics[scale=0.55]{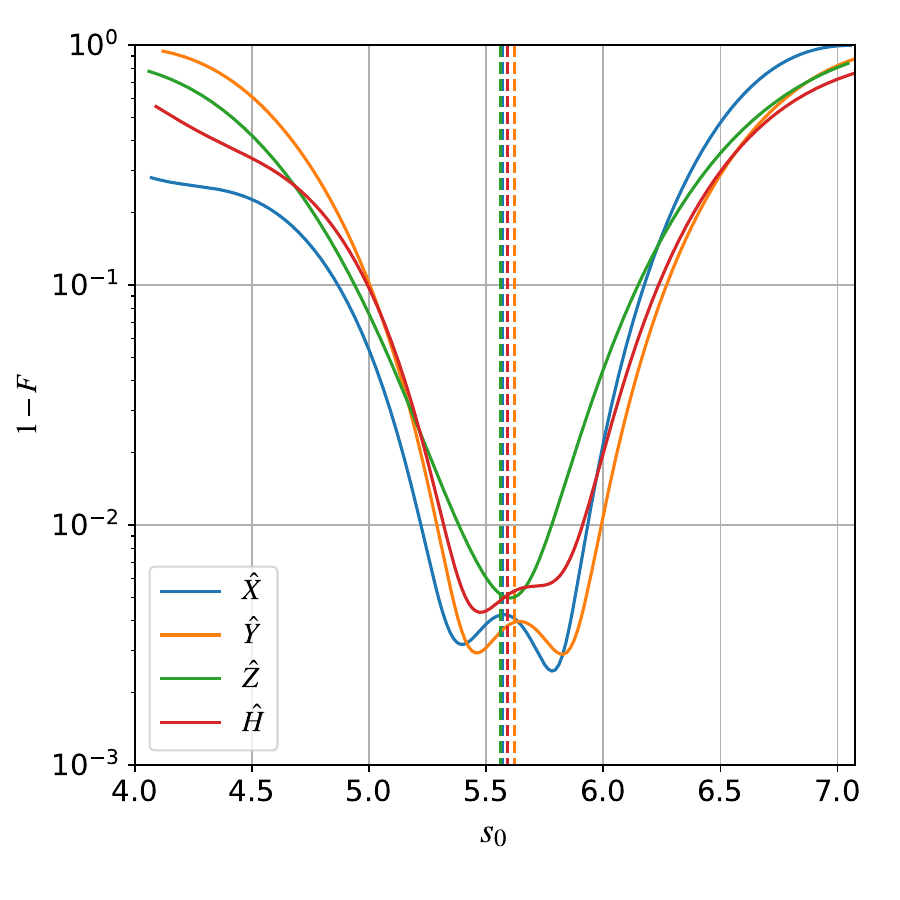}
    \caption{Numerical gate infidelities (see Eq. \eqref{Unitary_fidelity}) as a function of lattice depth after robust optimal control optimization. The colored dashed lines indicate the calibrated lattice depths before each gate experiment. }
    \label{qubit_gate_infidelity}
\end{figure}

\begin{table*}
\centering
\begin{tabular}{|c|c|c|c|c|c|c|c|c|}
\hline
\multirow{2}{*}{$\ \hat{U}_T\ $}& \multirow{2}{*}{$\epsilon_{exp}(\ket{0}\bra{0})$} & \multirow{2}{*}{$\epsilon_{exp}(\ket{1}\bra{1})$} & \multirow{2}{*}{$\epsilon_{exp}(\ket{+}\bra{+})$} & \multirow{2}{*}{$\epsilon_{exp}(\ket{-}\bra{-})$} & \multirow{2}{*}{$F_p(\epsilon_{th}, \epsilon_{oc})$} & \multirow{2}{*}{$F_p(\epsilon_{th}, \epsilon_{exp})$} & \multirow{2}{*}{$F_{ave}(\epsilon_{th}, \epsilon_{oc})$} & \multirow{2}{*}
{$F_{ave}(\epsilon_{th}, \epsilon_{exp})$}\\
& & & & & & & &\\
\hline

\multirow{4}{*} & & & & & & & &\\
{$\hat{X}$} & \multicolumn{1}{c|}{$P=0.991(6)$} & $P=0.978(15)$ & $P=0.981(15)$ & $P=0.962(14)$ & \multirow{2}{*}{$0.9958$} & \multirow{2}{*}{$0.880(17)$}&  \multirow{2}{*}{$0.9961$}  &  \multirow{2}{*}{$0.891(16)$}\\
 & \multicolumn{1}{c|}{$F_s=0.905(14)$} & $F_s=0.908(17)$ & $F_s=0.898(19)$ & $F_s=0.899(15)$ & & &  &  \\
 & & & & & & & &\\
\hline
\multirow{4}{*}
& & & & & & & &\\{$\hat{Y}$}& \multicolumn{1}{c|}{$P=0.806(30)$} & $P=0.849(28)$ & $P=0.813(56)$ & $P=0.795(34)$ & \multirow{2}{*}{$0.9960$} & \multirow{2}{*}{$0.801(24)$}& \multirow{2}{*}{$0.9962$}  & \multirow{2}{*}{$0.838(18)$} \\
 & \multicolumn{1}{c|}{$F_s=0.868(23)$} & $F_s=0.879(15)$ & $F_s=0.867(32)$ & $F_s=0.806(27)$ & & &  &  \\
 & & & & & & & &\\
\hline
\multirow{4}{*}
& & & & & & & &\\{$\hat{Z}$} & \multicolumn{1}{c|}{$P=0.903(49)$} & $P=0.909(43)$ & $P=0.981(7)$ & $P=0.863(27)$ & \multirow{2}{*}{$0.9948$} & \multirow{2}{*}{$0.887(16)$}&  \multirow{2}{*}{$0.9951$}  &  \multirow{2}{*}{$0.896(19)$}\\
 & \multicolumn{1}{c|}{$F_s=0.917(37)$} & $F_s=0.902(31)$ & $F_s=0.934(4)$ & $F_s=0.854(18)$ & & &   & \\
 & & & & & & & &\\
\hline
\multirow{4}{*}
& & & & & & & &\\{$\hat{H}$} & \multicolumn{1}{c|}{$P=0.845(20)$} & $P=0.886(14)$ & $P=0.799(17)$ & $P=0.778(28)$ & \multirow{2}{*}{$0.9949$} & \multirow{2}{*}{$0.889(9)$}&  \multirow{2}{*}{$0.9949$}  & \multirow{2}{*}{$0.893(8)$} \\
 & \multicolumn{1}{c|}{$F_s=0.859(19)$} & $F_s=0.903(11)$ & $F_s=0.836(13)$ & $F_s=0.829(27)$ & & &  & \\
 & & & & & & & &\\
\hline
\end{tabular}

\caption{\textit{SQPT results for qubit gates}. First column: target qubit gate $\hat{U}_T$. The next four columns show the purity $P$ and the quantum state fidelity $F_s$ between the ideally transformed input state and the experimental state. The next two columns present the process fidelity results, as defined by Eq.~\eqref{fid_2} between the ideal process and the process resulting from the OC procedure, and between the ideal process and the experimentally reconstructed one. The final two columns show the average gate fidelity of these processes, computed using Eq.~\eqref{general_fid_formula}. Experimental values and uncertainties are obtained from a bootstrap of the data. The four experiments were conducted sequentially, with the lattice depth calibration from one experiment setting the conditions for the next. Following the order of the rows in the table, the lattice depth calibration list is as follows: $s_{list}=[5.57, 5.62, 5.56, 5.59, 5.51]$.}
\label{tab:res_qubit}
\end{table*}

\begin{figure*}
    \centerline{
    \includegraphics[scale=0.4]{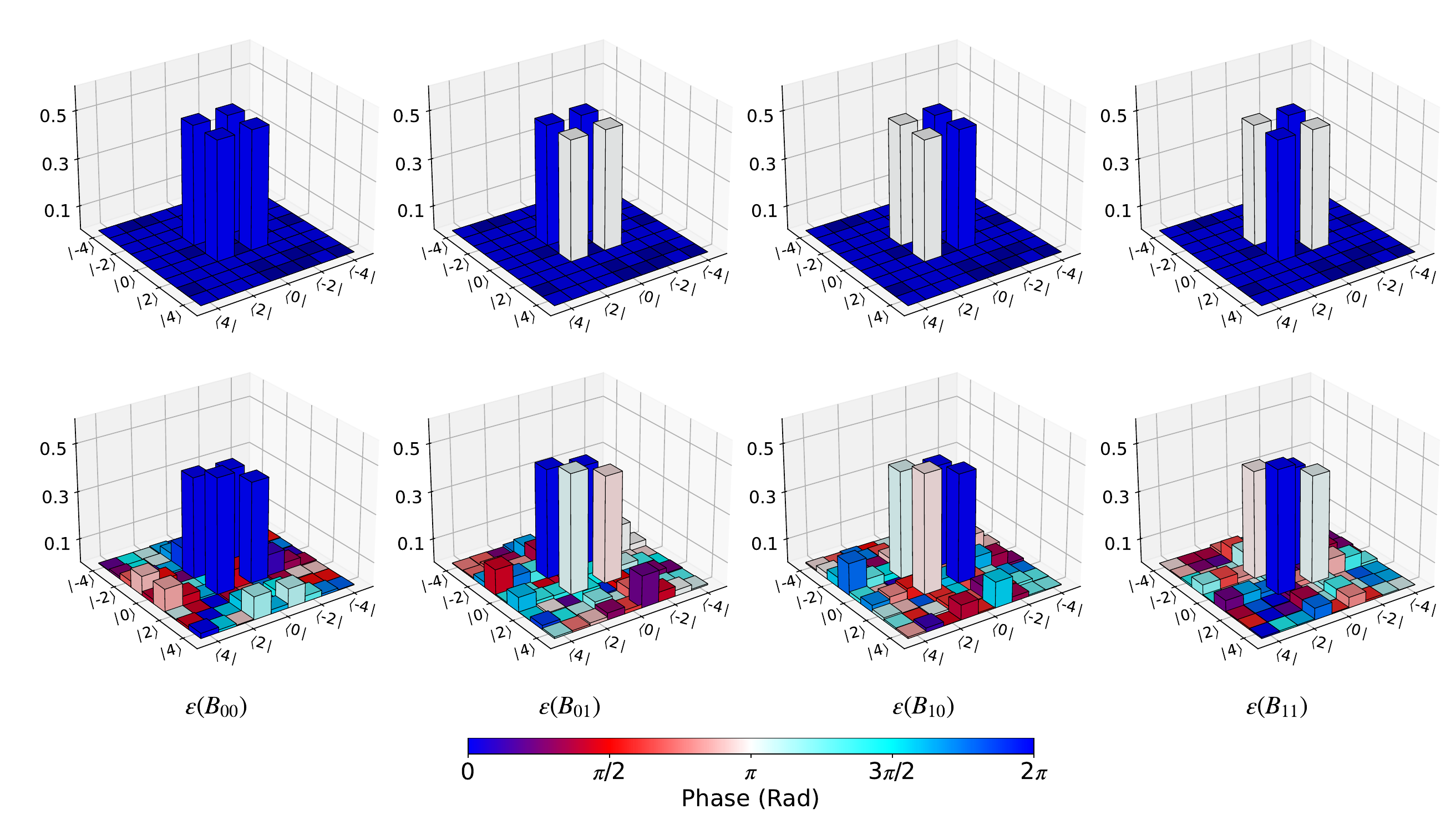}}
    \caption{Upper line (lower line): Representation of the submatrices composing the Choi matrix of the ideal Hadamard gate (experimental Hadamard gate) reconstructed using SQPT, in an extended $9$-dimensional momentum basis $\mathcal{H}'$ (See Sec.~\ref{SQPT_method}).}
    \label{représentation_hadamard}
\end{figure*}

For the $\hat{X}$, $\hat{Y}$, $\hat{Z}$ gates, and the Hadamard gate, the SQPT reconstruction basis is $\mathcal{B}=\{\ket{i}\bra{j}, i,j\in [-1, 1]\}$. Table \ref{tab:res_qubit} provides a summary of the experimental results. 
All input states were robustly prepared for the lattice depth set $\{s_0-0.5, s_0, s_0+0.5\}$ (control sequence duration~$=110$ $\upmu$s), and all gates were designed to maximize the fidelity over three lattice depth samples: $\{s_0-0.3, s_0, s_0+0.3\}$ (control sequence duration $=350$ $\upmu$s). The difference between state preparation and gates lattice depth samples comes from a tradeoff in order to limit the computational time of each optimization.

 Figure \ref{qubit_gate_infidelity} shows the gate infidelities as a function of the lattice depth.  The values of $s_0$ are given in Table \ref{tab:res_qubit}. We see a clear agreement between the reconstructed gates and the target gates. 
Imperfections in the reconstructed processes can be attributed to various factors, such as fluctuations in laser power, magnetic fields, or the number of atoms in the BEC. Other approximations, such as assuming a zero quasi-momentum width of the BEC or neglecting interactions, also lead to small discrepancies. Figure~\ref{représentation_hadamard} shows the ideal submatrices constituting the Choi matrix for the Hadamard gate in an extended $9\times9$ momentum basis $\mathcal{H'}$, alongside the experimentally reconstructed submatrices. As previously mentioned, all these results must be interpreted with caution, as the SQPT method intrinsically assumes perfect input states. Nevertheless, we obtain a very good approximation of the actual experimental process.

\begin{figure}[h!]
    \centering
    \includegraphics[scale=0.55]{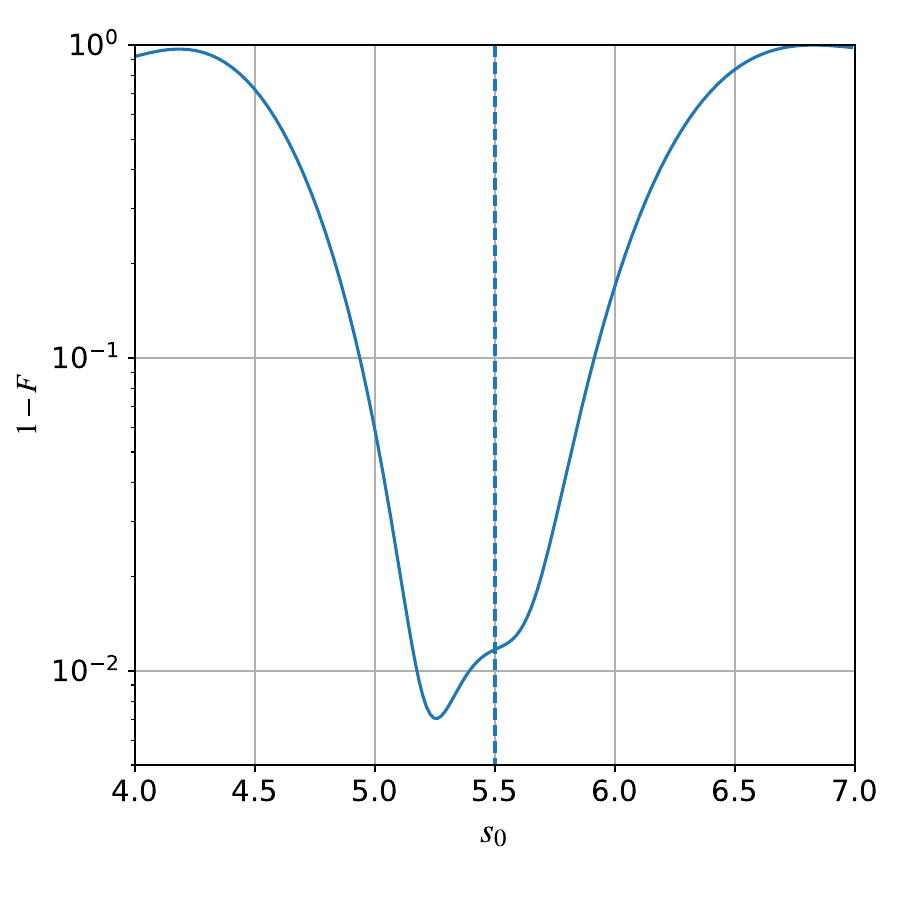}
    \caption{Numerical infidelity of the qutrit $\hat{X}^{(12)}$ gate (see Eq.~\eqref{Unitary_fidelity}) as a function of lattice depth. The dashed line indicates the calibrated lattice depth at the start of the experiment.}
    \label{infidelity_qutrit_gate}
\end{figure}

\subsection{Qutrit gate \label{qutrit_gate}}

For the qutrit $\hat{X}^{(12)}$ gate, we choose the symmetric momentum state basis $\mathcal{H}=\{\ket{-1},\ket{0}, \ket{1}\}$. The SQPT reconstruction basis is $\mathcal{B}=\{\ket{i}\bra{j}, i,j\in [-1,0, 1]\}$. All input states are prepared for the lattice depth set $\{s_0+(n/4)\times 0.8\}$, where $n$ is an integer varying from $-4$ to $+4$ (control sequence duration $= 110$ $\upmu$s).  The control for the qutrit gate was robustly designed using a set of seven lattice depths $\{s_0'+(n/3)\times 0.35\}$ where $n$ is an integer varying from $n=-3$ to $n=3$ (control sequence duration $=450$ $\upmu$s). This control sequence was optimized with a previous lattice depth calibration, so that $s_0'=5.39$ and $s_0=5.5$. Still, $s_0$ belongs to the robustness interval used in the control algorithm. Figure~\ref{infidelity_qutrit_gate} shows the numerical infidelity of the gate as a function of the lattice depth.\\
The average reconstruction fidelity for the nine elements of $\mathcal{B}$ is $\overline{F_s}=0.846(9)$ and the average purity is $\overline{P}=0.818(13)$, where the average is taken over the nine reconstructed states used for SQPT. Comparing the theoretical optimal control protocol and experimental quantum processes, we obtain the fidelities: $F_p(\epsilon_{th}, \epsilon_{oc}) = 0.9883$, $F_p(\epsilon_{th}, \epsilon_{exp}) = 0.820(12)$, $F_{avg}(\epsilon_{th}, \epsilon_{oc}) = 0.9896$ and $F_{avg}(\epsilon_{th}, \epsilon_{exp}) = 0.851(10)$.

It is worth noting that the fidelities of the reconstructed qutrit gate are comparable to those of the previous qubit gates, despite the fact that the control protocol requires significantly more time, as does the SQPT experiment itself. This highlights that the size of the controlled subspace is not the primary limitation in our ability to generate precise control protocols. Rather, the main limitation arises from experimental fluctuations that affect the quality of the reconstructed state during state tomography. 

\section{Applications}
\label{section5}
In this section, we aim to apply the evolution operator design to the stabilization of arbitrary qutrit states, as well as to the retrieval of the relative phase of a linear superposition of two quantum states.

\subsection{State stabilization}

\begin{table}[h!]
    \centering
    \begin{tabular}{|c|c|c|}
    \hline
    \multirow{3}{*}
    & & \\
    Input state & Preparation & Stabilization \\
        & & \\
        \hline
    \multirow{3}{*}
        & & \\
         $\ket{\psi}=\ket{-1}$ & $F_s=1$ &  $F_s=0.870(18)$ \\
         & & \\
         \hline
    \multirow{2}{*}
        & & \\
         $\ket{\psi}=\frac{\ket{0}+i\ket{1}}{\sqrt{2}}$ & \,$F_s=0.961(10)$\, & \,$F_s=0.869(21)$\, \\
        & & \\
                  \hline
    \end{tabular}
    \caption{Results from the identity operator applied to  the subspace $\mathcal{H}=\{\ket{-1},\ket{0}, \ket{1}\}$, applied to the states $\ket{-1}$ and $\frac{\ket{0}+i\ket{1}}{\sqrt{2}}$. The results for the first state have been averaged over three experimental realizations. The second state is reconstructed using a maximum likelihood estimation, with a bootstrap from eleven measurements.}
    \label{tab_identity}
\end{table}

As we can compute any desired unitary on a given momentum subspace, the identity operator is a valid target. This enables the generation of a stabilization operator without making assumptions about the input states, unlike previous works \cite{CRPHYS_2023__24_S3_173_0}, which either optimize state-to-state transitions or assume specific symmetries in the input states, thereby limiting stabilization to a single state. Here, we demonstrate an experimental realization of the identity operator on the subspace $\mathcal{H}=\{\ket{-1},\ket{0}, \ket{1}\}$, applied to the single momentum component $\ket{-1}$ and the superposition $\frac{\ket{0}+i\ket{1}}{\sqrt{2}}$. 
For the superposition state, we perform a full state tomography both before and after the gate to determine the fidelity to the desired superposition. This tomography also provides the purity of the state, which is $P=1$ for the preparation and $P=0.813(32)$ for the stabilization respectively.
For the single momentum state, the fidelity is straightforwardly measured from the population in state $\ket{-1}$.
The resulting fidelities are reported in Table~\ref{tab_identity}.

The unitary operation was numerically optimized to maximize fidelity over the values $\{s_0+(n/3)\times 0.4\}$ where $n$ is an integer varying from $n=-2$ to $n=2$ and $s_0=5.31$. The results demonstrate the effectiveness of this stabilization protocol, yielding state fidelities consistent with those presented in the previous sections.

\subsection{Phase retrieval}
\label{section5B}
Up to this point, all state tomographies have been performed using the maximum likelihood algorithm, by allowing the states to evolve in the static lattice and measuring the populations of momentum components during the evolution~\cite{dupont2023phase}. While this method is effective when the momentum components interfere sufficiently during evolution, it encounters limitations for certain classes of states, such as those involving distant momentum components in a shallow optical lattice. In such cases, the algorithm may return a strongly mixed state, as it cannot reliably detect the relative phase between the components. Increasing the lattice depth during the static evolution could enhance interference, but this approach might require very high laser power as well as a second lattice depth calibration before performing the state tomography.

Another approach to interference enhancement is to apply appropriate unitary operators.
In this section, we demonstrate how the protocols developed in this work can be used to retrieve the relative phase of a superposition of two momentum components. We consider two momentum subspaces, $\mathcal{H}_1=\{\ket{-1}, \ket{1}\}$ and $\mathcal{H}_2=\{\ket{-2}, \ket{2}\}$. To extract the relative phase in a superposition of two momentum components, we utilize the Hadamard gate and the operator $\hat{W}\equiv (\hat{Y}+\hat{Z})/\sqrt{2}$. A generic pure state in $\mathcal{H}_1$ or $\mathcal{H}_2$ can be expressed as $\ket{\psi}=a\ket{-j}+be^{i\theta}\ket{j}$ where $a^2+b^2=1$ and $j=1$ or $j=2$. 
The protocol requires three measurements. First, the coefficients $a$ and $b$ are determined by measuring the population in the momentum basis of the state $\ket{\psi}$. Next, by measuring the population of the states 
of $\hat{H}\ket{\psi}$ and  $\hat{W}\ket{\psi}$, the relative phase $\theta$ can be inferred since
\begin{eqnarray}
|\bra{-j}\hat{H}\ket{\psi}|^2  & = &  a^2+b^2+2ab \cos{(\theta)},  \nonumber \\
|\bra{j}\hat{H}\ket{\psi}|^2 & = & a^2+b^2-2ab \cos{(\theta)}, \nonumber \\
|\bra{-j}\hat{W}\ket{\psi}|^2 & = & a^2+b^2+2ab \sin{(\theta)}, \nonumber \\
|\bra{j}\hat{W}\ket{\psi}|^2 & = & a^2+b^2-2ab \sin{(\theta)}.
\label{sys_phase}
\end{eqnarray}

\begin{figure}
\centerline{
    \includegraphics[scale=0.55]{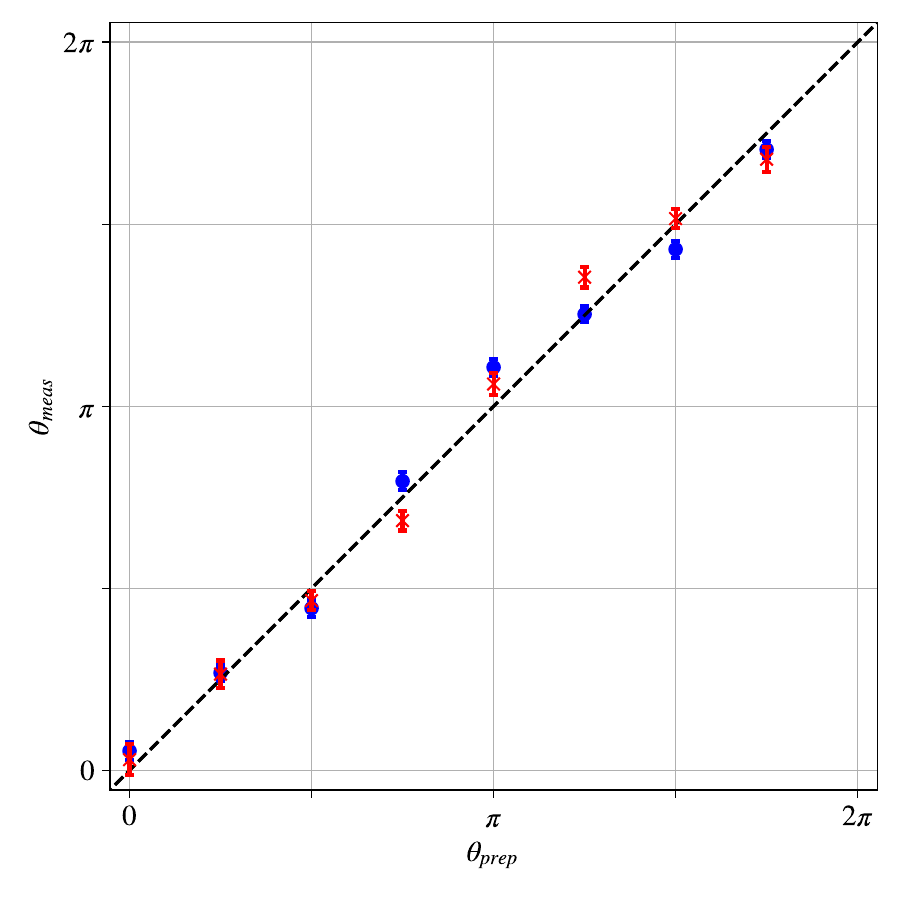}}
    \caption{Measured phases ($\theta_{meas}$) against various prepared phases ($\theta_{prep}$) of the state $\ket{\psi}=a\ket{-j}+be^{i\theta}\ket{j}$. The measurement was performed using the Hadamard gate and the $\hat{W}$ gate, with a least square algorithm applied to solve the resulting system given in \eqref{sys_phase} (see text).  The error bars indicate the $95\%$ confidence interval. Blue dots correspond to the basis $\mathcal{H}_1=\{\ket{-1}, \ket{1}\}$, while red crosses represent the basis $\mathcal{H}_2=\{\ket{-2}, \ket{2}\}$.}
    \label{figure_phase}
\end{figure}

A least squares algorithm can then be used to determine the most likely relative phase $\theta$, by comparing the measured populations to those of the ideally transformed state $\ket{\psi}$ with $\theta$ as the fitting parameter. Importantly, this method requires only three measurements, regardless of the lattice depth or the separation between the two components of the Hilbert space. We highlight this feature by applying the protocol to states in the momentum subspaces $\mathcal{H}_1$ and $\mathcal{H}_2$. We prepared eight states for each basis, with $a=b$ (numerical target), and phases $\theta_{prep} \in \{ 2\pi n/8\}$ where $n=0,1,...,7$, and measure the phase $\theta_{meas}$ using the protocol described above. Measurements were performed for a lattice depth of $s_0=5.26$ for the $\mathcal{H}_1$ Hilbert space and $s_0=5.03$ for $\mathcal{H}_2$. Figure \ref{figure_phase} displays the results of our experiments, revealing good agreement between the expected and measured phases. This underscores both the effectiveness of the proposed method in enhancing interference and, incidentally, the high quality of our quantum gates for this application.

\subsection{Extension and generalization of the method}

The method described in Sec.~\ref{section5B} is easily generalizable to arbitrary states in an $n-$dimensional Hilbert space. In fact, the $\hat{H}$ gate is nothing but the discrete Fourier transform for a qubit system. Similarly, the $\hat{W}$ gate is a modified Hadamard gate, where the upper (lower) diagonal is affected by a factor $-i$ ($i$). The generalization of our state estimation scheme is then straightforward. First, we measure the populations of the state to be determined. Second, we apply the Fourier transform to the state, by making sure that the dimension of the gate is equal or higher to the size of the state's Hilbert space, and measure the populations. Finally, we apply the modified Fourier Transform, where each upper diagonal element (lower) is affected by a factor $-i$ ($i$). This leads to a system of $2d$ equations, (where $d$ is the dimension of each gate) which constrains the relative phases of the state coefficients. This system is then numerically solved to retrieve the phases of each momentum component. To ensure that the solution found to the system of equations is optimal, it is solved a few hundred of times with different randomly generated initial conditions. This extra calculation is not a limitation, as it takes only a fraction of a second to solve the system once. Numerical simulation of this method succeeded in reconstructing a variety of randomly generated pure states in arbitrary high dimensional Hilbert spaces, so that experimental constraints would be the only limitation to the method.
An extensive implementation of our reconstruction scheme on arbitrary states is beyond the scope of this paper.

This use of controlled unitaries to induce interferences carrying information about the state can also be fruitfully combined with the maximum likelihood state reconstruction. This method, which we used to characterize our experimental gates, is a more general framework for state characterization, as it does not assume the state to be reconstructed to be a pure state, as is done in Sec.~\ref{section5B}.
The maximum likelihood algorithm only requires the knowledge of the transformation induced by the gates to yield a density-matrix state estimate. It is therefore straighforward to retrieve the phases between momentum components for the measurements of Fig.~\ref{figure_phase} from a maximum likelihood state estimation, which gives the same results, within uncertainties.
The purity of the estimated states however (which is no longer forced to 1) can be rather low, in particular for transformations in $\mathcal{H}_2$. Numerical studies show that this drop in purity can be attributed to a sensitivity of our control protocol to the quasi-momentum width of the finite size BEC.

It is possible to develop control protocols that are robust against the quasimomentum width, following the same procedure as in Sec.~\ref{alg} for the lattice depth. This was experimentally implemented with success in Ref.~\cite{Rodzinka:2024gtf}. 
Protocols robust against fluctuations of multiple physical quantities, \emph{e.g.} lattice depth and quasimomentum width, are achievable given enough control time, as the constraints are much stronger. Numerical simulations of such controls validate our methodology, and will be used in future works.

\section{Conclusion and perspectives}

In this article, we demonstrate the capability to generate arbitrary unitary transformations on a single cold atom qudit, utilizing the momentum components of a Bose-Einstein condensate (BEC) in a optical lattice as the computational basis. Our approach relies on a robust quantum control algorithm with the optical lattice position as the sole control parameter. Through full quantum process tomography of multiple quantum gates, we evaluate their quality using two fidelities widely accepted in the literature.

We experimentally demonstrate the versatility of our approach by generating a range of qubit operators  (such as $\hat{X}$, $\hat{Y}$, $\hat{Z}$, $\hat{H}$), and qutrit operators ($\hat{X}^{(12)}$, $\mathbb{I}_3$).  Using two qubit gates ($\hat{H}$, $\hat{W}$), we implement an efficient quantum state tomography scheme that enhances interference processes, enabling the extraction of relative phases between two momentum components in scenarios where previous methods failed. Our methodology represents a significant step forward in the context of high-fidelity qudit-based quantum computing and simulation \cite{PhysRevA.101.062307,Cao_2011,PhysRevC.108.064306, PhysRevLett.126.230504}. 
Our approach can readily be extended to multidimensional optical lattices, where additional degrees of freedom provide easier access to high-dimensionality qudits~\cite{dionis2025optimal}.
The tools developed here open up exciting possibilities for quantum sensing~\cite{PhysRevX.8.021059}, and in particular for fingerprinting methods~\cite{Ma2013,PhysRevA.96.053419}. 

\begin{acknowledgements}
This work was supported by the ANR projects QuCoBEC (ANR-22-CE47-0008), QUTISYM (ANR-23-PETQ-0002), and the EUR Grant NanoX (ANR-17-EURE-0009). D. G.-O. acknowledges support from the Institut Universitaire de France.
\end{acknowledgements}

\appendix*

\section{Proof for the average gate fidelity formula}\label{appendix}

We present a proof for the average gate fidelity used in our work. It is based on the following mathematical result~\cite{dankert2005efficientsimulationrandomquantum, PEDERSEN200747}:  
\begin{eqnarray}
\int d\psi &&\bra{\psi}M\ket{\psi}\bra{\psi}N\ket{\psi}= \nonumber\\
&&\frac{1}{d(d+1)}(\mbox{tr}(MN)+\mbox{tr}(M)\mbox{tr}(N)),
\label{Mathematical_result_non_trace_preserving_fid}
\end{eqnarray}

\noindent where $M,N$ are operators on a $d$-dimensional Hilbert space, and the integral is taken over the set of uniformly distributed pure states. 
The average gate fidelity 
defined in Eq.~\eqref{average_gate_fidelity} between a reconstructed process $\epsilon_{exp}$ and a target unitary process $\epsilon_{\hat{U}_T}$ can be reformulated as: 

\begin{equation}
    F_{avg}(\epsilon_{\hat{U}_T}, \epsilon_{exp}), =\int d\psi_i\bra{\psi_i}\hat{U}_T \epsilon_{exp}(\ket{\psi_i}\bra{\psi_i}))\hat{U}^{\dagger}_T \ket{\psi_i}.
\end{equation}
\noindent 
We consider $\epsilon_{exp}$  as a general quantum process written in terms of Kraus operators:
\begin{equation}
    \epsilon_{exp}(\rho)=\sum_{k}A_k \rho A_k^{\dagger}.
\end{equation}

\noindent Using Eq.~\eqref{Mathematical_result_non_trace_preserving_fid}, we obtain:

\begin{align}
     &F_{avg}(\epsilon_{\hat{U}_T}, \epsilon_{exp})=\sum_k\int d\psi_i\bra{\psi_i}\hat{U}_T A_k \ket{\psi_i}\bra{\psi_i} A_k \hat{U}_T^{\dagger} \ket{\psi_i} \nonumber \\
     &=\frac{1}{d(d+1)}\bigg(\mbox{tr}\bigg(\sum_k A_k^\dagger A_k\bigg) + \sum_k\abs{\mbox{tr}\big( U_TA_k^\dagger \big)}^2\bigg) \nonumber \\
     &=\frac{1}{d(d+1)}\bigg(\mbox{tr}(\epsilon_{exp}(\mathbb{I}_d))+d^2 F_p\big(\epsilon_{exp}, \epsilon_{\hat{U}_T}\big)\bigg),
     \label{demo_avg_fid}
\end{align}

\noindent where we used the fact that~\footnote{As for Eq.~\eqref{process_fidelity}, this result is obtained from the entanglement fidelity. (see Ref.\cite{Mele_2024})}: $F_p(\epsilon_{exp}, \epsilon_{\hat{U}_T})=\frac{1}{d^2}\sum_k\abs{\mbox{tr}\big( U_TA_k^\dagger \big)}^2$ (see Eq.~\eqref{process_fidelity}). The first term in this expression can be directly evaluated after SQPT since we have: 

\begin{equation}
    \alpha_{avg}\equiv \frac{1}{d}\mbox{tr}(\epsilon_{exp}(\mathbb{I}_d))=\frac{1}{d}\sum_i \mbox{tr}(\epsilon_{exp}(\ket{i}\bra{i})).
\end{equation}

\noindent Injecting $\alpha_{avg}$ in \ref{demo_avg_fid}, we finally obtain:

\begin{equation}
    F_{avg}(\epsilon_{exp}, \epsilon_{\hat{U}_T})=\frac{dF_p(\epsilon_{exp}, \epsilon_{\hat{U}_T}) + \alpha_{avg}}{d+1}, 
\end{equation}

\noindent which concludes the proof. 
 A different proof can be found in Ref.~\cite{PhysRevB.101.155306}, which follows the original derivation of Ref.~\cite{NIELSEN2002249}.

\bibliography{article.bib}

\end{document}